\documentclass{article}
\usepackage{enumitem}
\usepackage{amsmath}
\usepackage{amssymb}
\usepackage{physics}
\usepackage[final,dblblindworkshop]{tackling_climate_workshop_style}
\workshoptitle{Tackling Climate Change with Machine Learning}

\usepackage{graphicx}
\usepackage{pifont}
\usepackage{subcaption}
\usepackage{caption}
\usepackage{booktabs}
\usepackage{biblatex}
\renewbibmacro*{url+urldate}{%
  \usebibmacro{url}%
}

\usepackage{geometry}

\title{Unmasking Compensating Errors in Numerical Weather Models: A Process-Additive Neural Network Approach
  \thanks{DOI \protect\url{https://doi.org/10.48550/arXiv.2610.19772}.}}

\author{
  Yuiko Ichikawa$^{1,2,3*}$
  \And
  Edward Groot$^{2}$
  \And
  Jack W. Atkinson$^{4}$
  \AND
  \normalfont\small
  $^{1}$Environmental Change Institute, School of Geography and the Environment, University of Oxford\\
  $^{2}$Atmospheric, Oceanic and Planetary Physics, University of Oxford\\
  $^{3}$Department of Applied Mathematics and Theoretical Physics, University of Cambridge\\
  $^{4}$Institute of Computing for Climate Science, University of Cambridge\\
  $^{*}$Corresponding author: \texttt{yuiko.ichikawa@ouce.ox.ac.uk}
}
\begin{document}
\maketitle

\begin{abstract}
This work introduces a Process Additive Neural Network (PANN) framework to disentangle compensating biases in atmospheric models, where errors in different physical processes can offset each other and obscure their origins. Using a single-column model, each PANN branch represents the tendency error associated with a separate physical process, and the summed branch outputs approximate the total error relative to a reference tendency. Once trained successfully, the branch outputs are expected to provide process-level error attributions and, when averaged, estimates of process-level bias. Initial experiments focus on cloud-related processes in OpenIFS SCM case studies, where cloud and convection tendencies are strongly coupled and compensation is difficult to diagnose using standard approaches. The aim is not to replace human interpretation of physical processes, but to extend it to regimes where model complexity, process interaction, and compensating errors make direct diagnosis difficult. By providing a physically structured form of machine-learning interpretability, PANN can help connect complex model behaviour with process-level explanations. This work supports both improved understanding of compensating errors and the development of more physically consistent weather and climate models.
\end{abstract}

\section{Introduction}

This work applies a branched neural network architecture to identify and analyse \emph{compensating errors} in atmospheric models.

In many atmospheric systems, the total observed behaviour results from multiple distinct physical processes acting in superposition. While the combined outcome is observable, the contributions of individual processes are often hidden. This is particularly problematic when theoretical understanding of some processes is incomplete: errors in different processes can offset one another, making it difficult to attribute errors correctly, leading to a failure to understand the true nature of the processes. This phenomenon is known as \emph{compensating errors} \cite{Sandu.etal2020,Garcia-Diez.etal2015,Hourdin.etal2017}.

Numerical weather and climate models represent subgrid-scale processes---those unresolved by the model grid---through \emph{parameterizations}, simplified schemes that approximate their grid-box-mean effects. Among the most uncertain of these are cumulus convection (subgrid up- and downdraughts; \citep{Bechtold.etal2008}) and cloud microphysics (prognostic hydrometeors and large-scale precipitation; \citep{Forbes.etal2011}), often prototyped in \emph{single-column models} (SCMs) with prescribed lateral forcing, where process-level tendencies can be diagnosed before global trials.

Unfortunately, they are tightly coupled, and their humidity tendencies are strongly anticorrelated, as convective detrainment feeds the cloud scheme \citep{ECMWF2023}. Standard post-hoc interpretability methods, such as Shapley Additive Explanations (SHAP), input-gradient saliency maps \citep{Simonyan.etal2013}, or partial radiative perturbation \citep{Wetherald.Manabe1988}, struggle in the presence of such high autocorrelation, leading to high variance in attributions and potential ``disagreement'' between explanation methods \citep{Ray-Mukherjee.etal2014,Aas.etal2021,Verdinelli.Wasserman2023,Flora.etal2022,Krell.etal2025,Heuer.etal2024}.
Krell et al.\ and Flora et al.\ showed that strong correlations among input features can destabilize individual-feature attributions, and both advocate grouping correlated variables to obtain more robust and interpretable explanations \citep{Krell.etal2023,Flora.etal2024,Krell.etal2025}. While such grouping provides a remedy, our PANN approach addresses the problem architecturally by enforcing a parallel structure that mirrors the additive decomposition of physical processes in Numerical Weather Prediction (NWP). This structure constrains the model to represent process contributions separately, in a form more closely aligned with the physical organization of the underlying system.

We represent the model as a Process Additive Neural Network, in which each branch corresponds to a distinct physical process and learns its own hidden representation. The outputs of all branches are \emph{summed} to produce the overall prediction. These hidden representations can be analysed post-training to identify correspondence with physical variability and potential error structures.

\section{Method}
\paragraph{Architecture of the Process Additive Neural Network and learning problem.}
The architecture follows the Neural Additive Model \citep{Agarwal.etal2020}: one branch per process, with no cross-branch connections in the hidden layers (Fig.~\ref{fig:PANN_schematics}). The details of each blocks are described in the Supplement. PANN is trained to attribute the error $\hat{e}$ in specific-humidity tendency $dQ$ to individual physical processes. Each branch receives only the tendency of one process: $dQ_{val}$. Branch outputs: $\hat{e}_{val}$ are summed, and that sum is trained to match the discrepancy between a verification error $e_{truth}$ (with \emph{no} process breakdown) and the predicted error $\hat{e}$. Individual process errors are never observed; they are inferred as $\hat{e}_{val}$ from this additive constraint.

This design has three practical roles. First, it turns an additive physical constraint into an attribution mechanism: one observed total is decomposed into process contributions. Second, connections that the physics does not allow are removed \emph{a priori}, so the hidden layers cannot mix convection with cloud. Third, the model remains simple enough that standard XAI tools (e.g. SHAP or LOCO) apply with little change; because the branches do not mix, those tools need only consider combinations of a few process heads rather than of all hidden units.

\begin{figure}[t]
\centering
\includegraphics[width=0.9\linewidth]{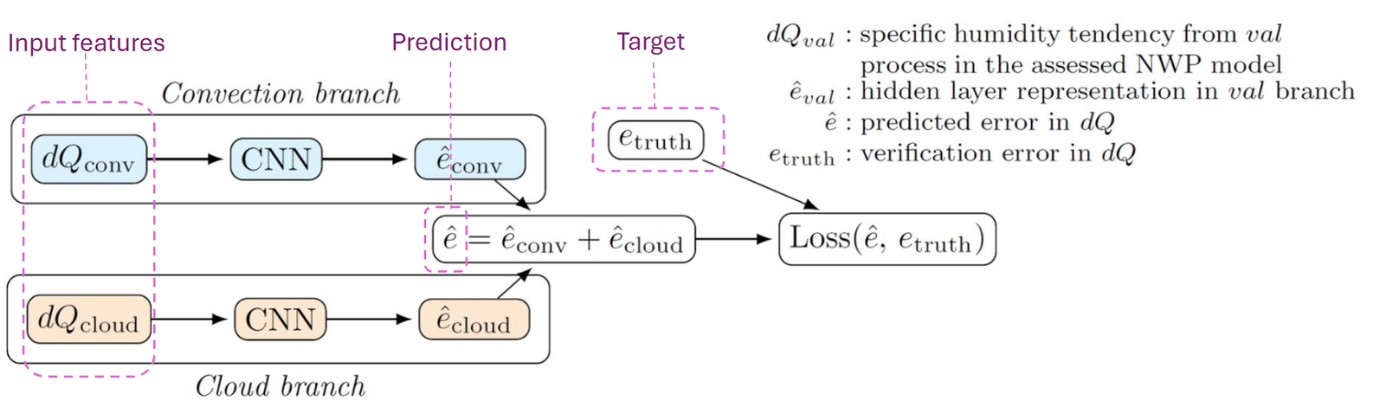}
\caption{Schematic of the Process Additive Neural Network (PANN) with two process branches.}
\label{fig:PANN_schematics}
\end{figure}

\paragraph{The NWP model.}
The model under test is ECMWF OpenIFS cycle 48r1 run as an SCM. OpenIFS physics comprises eight parameterization branches, but specific-humidity tendency $dQ$ is dominated by cloud and convection\footnote{In OpenIFS Cycle 48r1 onward, the turbulent boundary-layer scheme produces no $dQ$: its mass flux is restricted to clear-sky columns, with all moist vertical transport and cloud-related moisture tendencies assigned to the convection and prognostic cloud schemes. Vertical diffusion, the only other parameterization contributing to $dQ$, has a much smaller impact.}. We therefore restrict PANN to these two branches. Both are highly uncertain, and their humidity tendencies are strongly anticorrelated, making this pair a stringent test of compensating errors while remaining faithful to the model's humidity budget.

\paragraph{Reference data and experimental stages.}
Forcing and reference states come from Christensen et al.\ \cite{christensenForcingSingleColumnModels2018} (C18): convection-permitting 4\,km MetUM Cascade simulations, coarse-grained to drive the parameterized OpenIFS SCM, over the tropical oceans ($20^\circ$S--$20^\circ$N, $42^\circ$E--$177^\circ$E). Prescribed humidity advection is removed so both supply physics-only $dQ$. We draw $\sim$10{,}000 independent columns at random from the domain and retain 15-min snapshots of $dQ$.

Parameterization development usually inspects mean error or \emph{bias} after averaging over many steps ---often six hours or longer---whereas PANN is trained on single-step discrepancies. That average is convenient for human diagnosis, but it suppresses the step-to-step fluctuations that identify the responsible process. Humans prefer averaging to simplify the problem; a learning model can instead learn from the noisy residual, comparing steps with active cloud but weak convection to steps with both branches active to separate cloud from convection despite their anticorrelation. We therefore train on step errors and recover bias afterwards as their average. The same short time step also makes the method applicable to field-campaign datasets, which typically offer limited spatial coverage and duration (e.g. TWPICE \cite{May.etal2008}).

Two stages are used. Stage~1 plants a synthetic total error with \emph{known} cloud and convection contributions and tests whether PANN recovers the prescribed branch contributions. Stage~2 applies the same architecture and training to the OpenIFS--C18 $dQ$ difference, where per-process truth is unavailable. Sampling, time discretisation, and the synthetic target formula are given in the Supplement.

\section{Estimated error attributions}
\subsection{Synthetic error}

\begin{figure}[t]
  \centering
  \begin{minipage}[c]{0.75\linewidth}
    \centering
    \includegraphics[width=\linewidth]{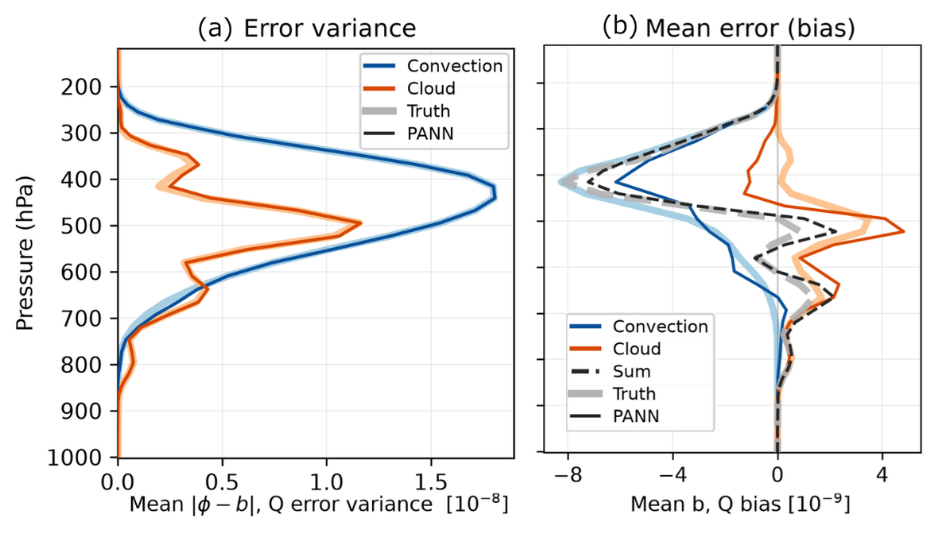}
  \end{minipage}\hfill
  \begin{minipage}[c]{0.25\linewidth}
    \caption{Branch attribution of the synthetic error; truth in thick pale, PANN in thin saturated, convection blue and cloud orange. (a) Mean absolute importance. (b) Mean error, sum dashed.}
    \label{fig:synthetic}
  \end{minipage}
\end{figure}

Stage~1 asks a controlled question: when the cloud and convection components of a total error are known, can PANN recover them separately? The synthetic target (Supplement) mimics compensating biases---upper-tropospheric convective moistening against mid-level cloud drying---before the method is applied to real SCM discrepancies. Attribution of error variance is nearly exact, whereas attribution of the mean error (bias) is less identifiable (Fig.~\ref{fig:synthetic}). Branch error variances are largest in the mid-troposphere, where convection and cloud errors are simultaneously active, providing the strongest test of separation. PANN successfully predicts the total error variance and attributes it to the individual branches (Fig.~\ref{fig:synthetic}a). In Fig.~\ref{fig:synthetic}b, PANN recovers the planted branch-level biases with small discrepancies in upper levels, where it may assign a weak negative cloud bias despite the planted cloud bias being negligible. Mean error (bias) attribution is intrinsically harder because it does not depend on individual inputs: constant offsets can be redistributed between branches during training without affecting the result, remaining an identifiability problem.

\subsection{Error against the C18 cloud-permitting reference}

\begin{figure}[t]
  \centering
  \includegraphics[width=1.0\linewidth]{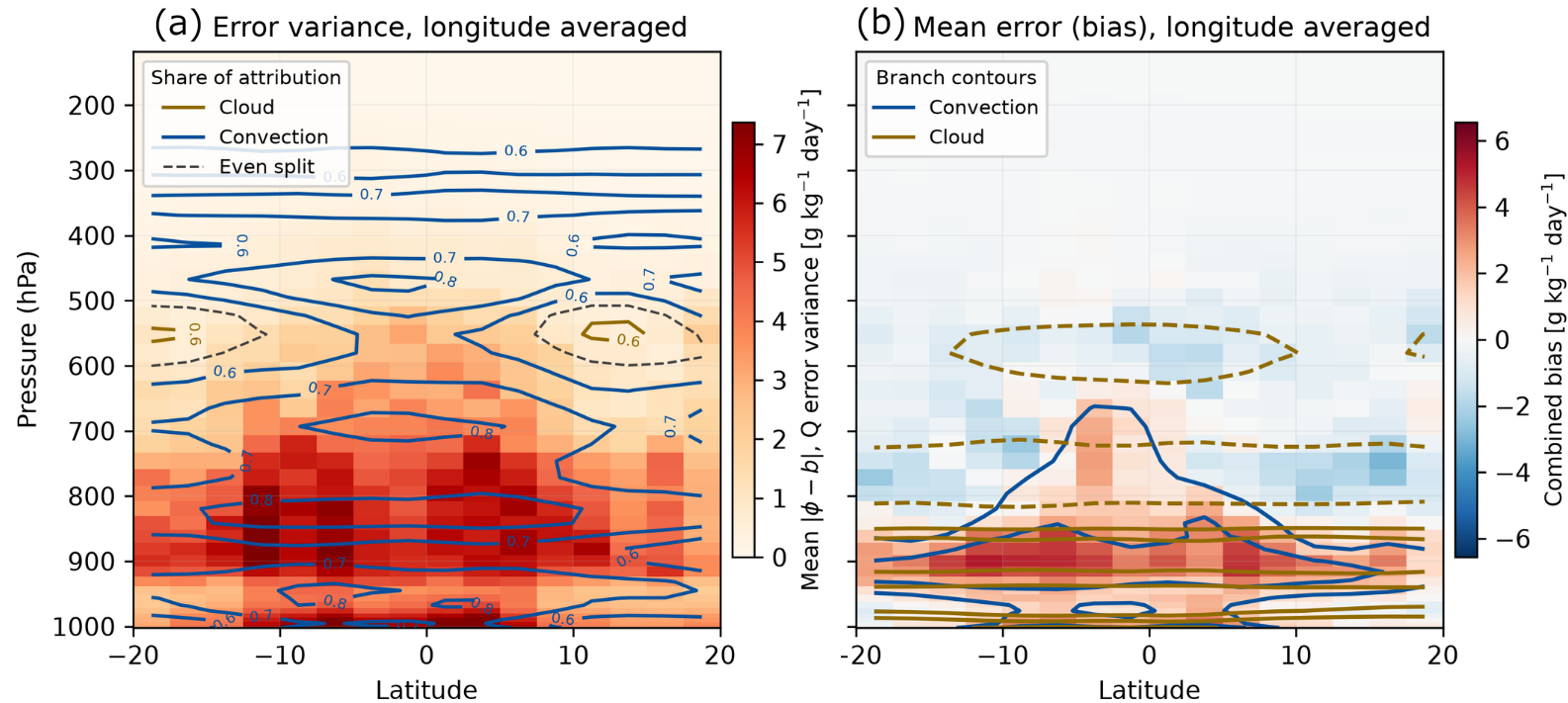}
  \caption{Branch attribution of specific humidity tendency discrepancy between C18 and OpenIFS, longitude-averaged over the tropical Indo-Pacific ocean. (a)~Variance of $dQ$ error (shading) with the convection (blue) and cloud (khaki) share of branch amplitude contoured; dashed grey is the 50/50 split. (b)~Mean $dQ$ bias (shading) with branch contributions contoured (convection blue; cloud khaki for solid/dashed positive/negative).}
  \label{fig:ukmo}
\end{figure}

Having validated PANN on synthetic targets, we apply it to the $dQ$ difference between OpenIFS and C18 (Fig.~\ref{fig:ukmo}). Panel (a) shows that error variance is largest in the lower troposphere (surface--$\sim$700\,hPa) and enhanced in the equatorial convection belt, where moisture variability and cloud-convection activity are highest. Convection dominates the branch-amplitude share over most of the section, but mid-tropospheric pockets ($\sim$500--650\,hPa, $\sim$10--15$^\circ$ from the equator) have a larger cloud share. Panel (b) shows a different structure. The mean signed bias is strongly positive in the boundary layer, contributed by both branches. Aloft, cloud contributions turn negative while convection remains positive, so the combined mean bias is much smaller---a partial cancellation between the two moist branches that variance alone obscures. The height-dependent sign change in the cloud branch may reflect microphysical transitions (e.g.\ around the freezing level) and warrants comparison with scheme thresholds. Crucially, the convection bias is capped by a dome-shaped boundary peaking near 700\,hPa at the equator and sloping down to 900\,hPa at $\pm20^\circ$. This traces the convective boundary layer and may reflect moisture detrainment from shallow convection.
The contrast between the variance and mean-error patterns is informative. The error variance primarily reflects spatial and temporal mismatches in the occurrence of active convective/cloud grid points between the convection-explicit C18 reference and parameterized OpenIFS. In contrast, the signed mean error reveals process-specific bias structures that may be associated with structural deficiencies in the parameterizations. Importantly, the error variance plays a key role in the PPNN learning process: it helps constrain the estimation of mean-error (bias) parameters, whose correct attribution is more challenging, by binding them to the pointwise on–off mismatches.

\section{Discussion}
An important question is whether these attributions can be trusted when the true process-level errors are unknown. The answer is necessarily qualified. In the synthetic experiment, where the truth is prescribed, PANN recovers the attribution well. For the C18 experiment, however, the branch contributions should be interpreted as a physically constrained diagnosis rather than as verified truth. They identify a decomposition that is consistent with the observed total tendency discrepancy and the specified process structure, but they cannot by themselves prove the underlying causal error in either parameterization.

PANN is not intended to replace physical understanding or expert judgement. Instead, it extends process-based interpretation to problems in which several parameterizations act simultaneously and their errors compensate. This capability will become increasingly important as atmospheric models grow more complex. Owing to limits in observations and computational cost, models have historically represented complex systems through simplified, aggregated parameterizations while retaining their essential behaviour. As computational resources increase, models may represent more of this complexity explicitly; diagnosing the resulting interactions will remain challenging. Methods such as PANN can help reveal patterns arising from interacting processes without requiring those patterns to be reduced to a simpler system first.

This study demonstrates the approach only for two branches and for specific humidity. Temperature tendencies depend on more physical processes---approximately eight parameterizations contribute to the relevant OpenIFS tendencies---and therefore pose a more difficult attribution problem. Nevertheless, cloud-related processes remain among the largest sources of uncertainty. A humidity attribution can consequently provide useful constraints and hypotheses when extending the method to temperature, particularly if the cloud-error representation is held fixed or otherwise informed by the humidity analysis (e.g. hierarchical calibration in \cite{Varejao.etal2013,Kang.etal2015}).

The reference simulation also requires careful interpretation. The C18 cloud-permitting MetUM has its own structural and parameterization biases; higher spatial resolution does not automatically make it closer to the truth. Physics suites forced by the same dynamics can also agree with each other while remaining distinct from a convection-permitting benchmark \cite{groot_how_2026,Groot.etal2026}. A valuable next step would therefore be to apply the same workflow to field observations, for example from TWPICE \cite{May.etal2008}. This would test whether PANN can diagnose model errors against observations, rather than only differences between two models.

\section{Impact}
PANN offers a process-aware way to diagnose compensating errors in weather and climate models. Parameterization workflows rely heavily on calibration against bulk metrics, which need not expose competing biases between schemes \cite{Hourdin.etal2017,Garcia-Diez.etal2015}; PANN can diagnose when such compensation occurs and indicate which parameterizations---and, in regional applications, which locations---should be prioritised for revision. By separating a total tendency discrepancy into process-level contributions, it can guide tuning towards physically consistent fixes rather than net-bias reduction alone. Because iterative calibration requires many costly model integrations to explore uncertain parameters \cite{Hourdin.etal2017} and is a major bottleneck in parameterization development \cite{Couvreux.etal2021}, better-targeted diagnostics could make that loop more efficient. In the longer term, this may improve our understanding of the atmospheric processes represented by these schemes and support more reliable weather and climate simulations.

\begin{ack}
This work was supported by an Environmental Change Institute Small Grant, University of Oxford, and the Alan \& Christina MacDonald Award from Queens' College, Cambridge.

Y.I. designed and implemented the method, performed the experiments and analysis, and wrote the manuscript. E.G. and J.A. advised on the implementation, reviewed the manuscript, and contributed to refining the work.

We thank Evan Krell, Chris Edsall, Yohei Takano, Sebastian Schemm, Lorena Escudero Sánchez, members of the AOPP Predictability of Weather and Climate Group at the University of Oxford, and three anonymous reviewers for their valuable comments and advices.
\end{ack}

\clearpage
\appendix

\section*{Supplementary Materials}

\setcounter{section}{0}
\renewcommand{\thesection}{S\arabic{section}}

\section{PANN architecture and training cycle}
\label{sec:supp-pann}

\subsection{Network structure}

The Process Additive Neural Network (PANN) assigns one independent sub-network
(\emph{branch}) to each physical-process partition.  Each branch maps its
process-specific input $dQ_{val}$ to a vertical profile of predicted error
$\hat{e}_{val}$; branch outputs are summed to give the total predicted error
\begin{equation}
  \hat{e} = \sum_{p=1}^{P} w_p \, \phi_p(dQ_{val,p}),
  \label{eq:pann-sum}
\end{equation}
where $\phi_p$ is branch~$p$, $w_p$ is a non-negative importance weight with
$\sum_p w_p = P$ under the default normalisation, and only the summed
prediction $\hat{e}$ is supervised against $e_{\mathrm{truth}}$.

\paragraph{Branch inputs and outputs.}
We use two branches---convection and cloud microphysics---each receiving the
OpenIFS specific-humidity tendency $dQ_{val}$ for that process ($P=2$,
$Q$-only output).  Each branch outputs one value per retained model level
($n_{\mathrm{out}}=1$).

The runs use the IFS 137-level grid (L137), but training and evaluation exclude
the mid-upper stratosphere (92th level and above), where specific-humidity increments are small and
noisy and hinder the more important tropospheric/lower stratospheric attribution.

\paragraph{Branch architecture.}
Each branch is a multi-scale one-dimensional convolutional network: three
parallel Conv1d paths (2 paths with kernel sizes $k=3$ and 1 path with $k=4$ with dilation by $k=1$) capture local spikes and their shifts; their outputs are concatenated, merged, and compressed to a 10-channel vertical-mode bottleneck before a symmetric decoder with additive skip connections reconstructs the profile. ReLU activations are used
throughout.

\paragraph{Level blocking.}
Adjacent model levels carry correlated error structure, so the column is split into overlapping vertical blocks (four levels, two-level overlap), trained independently and stitched at evaluation with linear blending in the overlap zones.

\subsection{Training cycle}

Training follows a standard supervised workflow with two PANN-specific steps.

First, a short \emph{dominance probe} on the initial cross-validation fold
estimates fixed branch weights.  Under leave-one-covariate-out (LOCO) ordering,
the importance of branch~$p$ is the drop in validation $R^2$ when that branch
is removed:
\begin{equation}
  I_p = R^2(\text{all branches}) - R^2(\text{all branches} \setminus p).
  \label{eq:loco}
\end{equation}
LOCO down-weights branches that correlate with the target but do not improve
skill when removed.  Weights $w_p = \max(I_p, 0)$ are renormalised so that
$\sum_p w_p = P$.

Second, the model is trained by $k$-fold cross-validation with inputs
standardised on the training fold.  Optimisation uses Adam with early stopping;
the objective is mean squared error weighted by $p/p_{\mathrm{sfc}}$ at each
level.  Per-branch contributions are the anchored and weighted branch outputs of
Eq.~\eqref{eq:anchor}.

\subsection{Anchored branch contributions}
\label{sec:supp-anchor}

Because only the sum in Eq.~\eqref{eq:pann-sum} is supervised, the branch
outputs are identifiable only up to constants: adding constant bias terms $\mathbb{E}[\phi_p]$ to branches
such that they cancel each other $\sum_p w_p\mathbb{E}[\phi_p] = 0$ leaves every prediction
unchanged. Due to this, the mean
contribution $\mathbb{E}[w_p\phi_p]$ is more difficult, such that the split of the mean error
(bias) between branches is set by the initialisation and optimisation path
rather than by the data.  Input standardisation adds to this, because an
inactive process ($dQ_{val,p} = 0$) does not map to a zero network input.

We remove this freedom by anchoring every branch at its physically inactive
input:
\begin{equation}
  \hat{e} = \hat{e}_0 + \sum_{p=1}^{P} \hat{e}_p, \qquad
  \hat{e}_p = w_p\left[\phi_p(dQ_{val,p}) - \phi_p(\mathbf{0})\right], \qquad
  \hat{e}_0 = \sum_{p=1}^{P} w_p\,\phi_p(\mathbf{0}),
  \label{eq:anchor}
\end{equation}
where $\mathbf{0}$ denotes a zero tendency profile (after standardisation)
and $\hat{e}_0$ is an intercept not attributed to any process; in physical
units it also absorbs the target mean removed by standardisation.
Eq.~\eqref{eq:anchor} is an exact rewrite of Eq.~\eqref{eq:pann-sum}, so
predictions and skill are unchanged and anchoring is applied to trained
models without retraining.  It fixes the constants with a physical
convention: a process that is inactive contributes no error,
$\hat{e}_p = 0$ when $dQ_{val,p} = 0$.

This is the anchored form of a
functional decomposition, with the anchor at the inactive state rather than
at the data mean \cite{kuo_decompositions_2009}.  For an additive model its structure and thinking coincides
with Baseline Shapley attribution relative to that state
\cite{sundararajan_many_2019}.  When the planted error vanishes for
inactive processes, as in Section~\ref{sec:supp-synthetic}, the recovered
$\hat{e}_0$ should be close to zero.

\section{Construction of the tendency dataset}
\label{sec:supp-dataset}

Forcing, boundary conditions, and reference states come from Christensen et al.\ \cite{christensenForcingSingleColumnModels2018} (C18):
convection-permitting 4\,km MetUM simulations over the tropical Indo-Pacific
($20^\circ$S--$20^\circ$N, $42^\circ$E--$177^\circ$E), coarse-grained to drive
the OpenIFS~48r1 SCM.  Stage~2 therefore compares parameterised OpenIFS moist
physics with resolved MetUM moist processes.

The dataset comprises $\sim$10{,}000 independent columns drawn at random from
the domain, taken from pre-computed C18 and SCM runs with 15-minute time
stepping.  Prescribed humidity advection is removed so both models supply
physics-only $dQ$.  Architecture and training are in
Section~\ref{sec:supp-pann}; the Stage~1 planted target is in
Section~\ref{sec:supp-synthetic}.

\paragraph{Stage~2 target.}
The training target is the specific-humidity tendency discrepancy between C18
and OpenIFS:
\begin{equation}
  e_{\mathrm{truth}} = dQ_{\mathrm{C18}} - \overline{\mathrm{qadv}}
      - dQ_{\mathrm{total}},
  \label{eq:ceda-dq}
\end{equation}
where $dQ_{\mathrm{C18}}$ is the C18 physics tendency,
$dQ_{\mathrm{total}}$ the OpenIFS total moist-physics tendency, and
$\overline{\mathrm{qadv}}$ the mean prescribed advective moisture forcing over
the sample window.  Subtracting $\overline{\mathrm{qadv}}$ removes shared
large-scale moisture transport so that disagreements reflect parameterisation
differences rather than boundary drift.

\section{Synthetic error formulation}
\label{sec:supp-synthetic}

Before applying PANN to real SCM errors, we validate the architecture with a
planted $e_{\mathrm{truth}}$ for which per-process contributions are known.
Both branches receive their respective OpenIFS $dQ_{val}$ profiles as inputs.

\subsection{Design intent}

The synthetic target mimics a compensating humidity-error pair: convection moistens too strongly aloft while cloud processes dry too strongly at mid-levels.  Those errors can cancel in the column integral while remaining large level by level---the identifiability problem PANN is meant to address.

\subsection{Height-dependent Gaussian target}
\label{sec:supp-linear}

Let $k$ denote the model-level index.  Gaussian envelopes centred on the level
nearest pressure $p_\star$ with width $\sigma_k$ (in level units) are
\begin{equation}
  E(k;\, p_\star, \sigma_k)
  = \exp\!\left[-\tfrac{1}{2}\left(\frac{k - k(p_\star)}{\sigma_k}\right)^2\right].
\end{equation}
The planted error is
\begin{equation}
  e_{\mathrm{truth}}(k) = \alpha_{\mathrm{conv}}\, E(k;\, p_{\mathrm{conv}}, \sigma_{\mathrm{conv}})\,
           dQ_{\mathrm{conv}}(k)
       + \alpha_{\mathrm{cloud}}\, \widetilde{E}(k;\, p_{\mathrm{cloud}}, \sigma_{\mathrm{cloud}})\,
           dQ_{\mathrm{cloud}}(k),
  \label{eq:gaussian-error}
\end{equation}
where the cloud envelope includes vertical \emph{zig-zag} modulation
\begin{equation}
  \widetilde{E} = \frac{E \cdot \left[1 + A\cos\!\left(2\pi(k-k_\star)/P\right)\right]}{1+A},
\end{equation}
with ripple amplitude $A$ and period $P$ (in levels), so cancellation is
incomplete level by level even when the column integral is small.

\paragraph{Parameter values.}
$\alpha_{\mathrm{conv}}=1.1$, $p_{\mathrm{conv}}=400$~hPa,
$\sigma_{\mathrm{conv}}=4$ levels;
$\alpha_{\mathrm{cloud}}=0.9$, $p_{\mathrm{cloud}}=500$~hPa,
$\sigma_{\mathrm{cloud}}=5$ levels;
$A=0.6$, $P=6$ levels.
This yields upper/mid-tropospheric convective moistening and mid-level cloud
drying with vertical ripple---a compensating pair that cancels more in the
column integral than level-by-level.

\paragraph{Evaluation against truth.}
Known per-branch contributions are compared with PANN branch attributions.
Mean absolute importance and branch share of total importance test variance
attribution; mean error profiles test recovery of the signed, time-mean
components given that the planted coefficients are input-independent.

\subsection{Non-linear (Chebyshev) planted error}
\label{sec:supp-chebyshev}

To test whether attribution survives a non-linear error--tendency
relationship, the linear factor $dQ_p$ in Eq.~\eqref{eq:gaussian-error} is
replaced by a fourth-degree Chebyshev response, with the envelopes unchanged:
\begin{equation}
  g_p\!\left(dQ_p\right) = s_p(k) \sum_{n=1}^{4} c_{n,p}
      \left[T_n(u) - T_n(0)\right],
  \qquad
  u = \operatorname{clip}\!\left(\frac{dQ_p(k)}{s_p(k)},\, -1,\, 1\right),
  \label{eq:chebyshev-error}
\end{equation}
where $T_n$ is the Chebyshev polynomial of degree $n$ and $s_p(k)$ is the
99th percentile of $|dQ_p|$ at level~$k$.  The scale maps tendencies onto
the natural domain $[-1, 1]$ of $T_n$, and clipping makes the most extreme
1\% saturate instead of growing as $u^4$.  Subtracting $T_n(0)$, which is
non-zero for even $n$, keeps $g_p(0) = 0$.  An inactive process therefore
still contributes no error.  Without this, each branch would carry a
constant that is not identifiable even in the planted truth.  With
$c_{\cdot,p} = (1, 0, 0, 0)$ the response reduces to $dQ_p$ within the clip
range, which recovers Eq.~\eqref{eq:gaussian-error}.  We use
$c_{\cdot,\mathrm{conv}} = (1.0, 0.5, -0.3, 0.4)$ and
$c_{\cdot,\mathrm{cloud}} = (1.0, -0.5, 0.3, -0.4)$.  Both responses are
strongly curved and non-monotonic on $[-1, 1]$: their correlation with the
linear planted error is only 0.36 (convection) and 0.64 (cloud).  Their even
terms also shift the planted mean error, so the bias attribution is not
inherited from the linear case.  All other settings are identical to the linear experiment.

\begin{figure}[h]
  \centering
  \includegraphics[width=\textwidth]{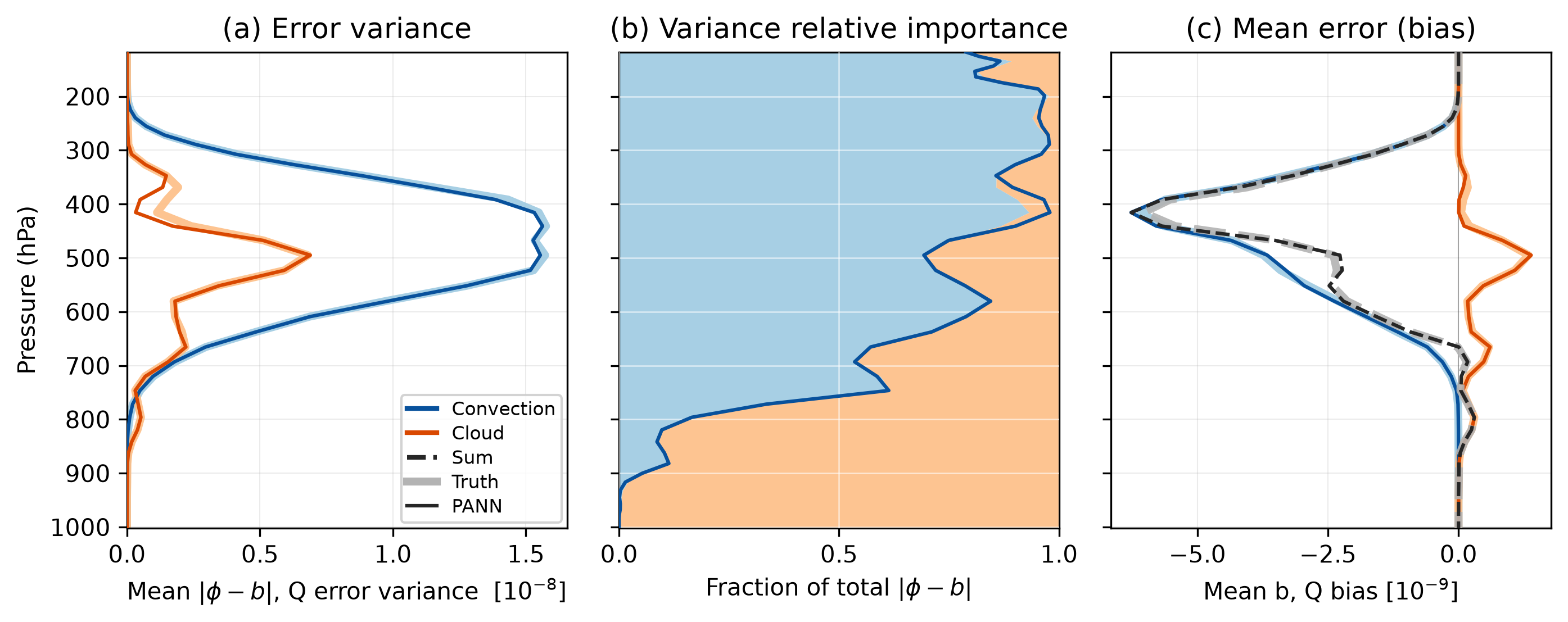}
  \caption{Branch attribution for the Chebyshev planted error (Eq.~\eqref{eq:chebyshev-error}), as in Fig.~\ref{fig:synthetic}: thick pale, truth; thin saturated, PANN (convection blue, cloud orange). (a) Mean absolute importance. (b) Branch share of total importance. (c) Anchored mean error (bias) by branch; sum dashed.}
  \label{fig:synthetic-chebyshev}
\end{figure}

\paragraph{Synthetic error attributions (continued).}
The main text reports attribution for the linear target (Fig.~\ref{fig:synthetic}); Section~\ref{sec:supp-linear} gives the target definition.
PANN recovers both linear and non-linear planted errors (Figs.~\ref{fig:synthetic} and~\ref{fig:synthetic-chebyshev}).
For the Chebyshev target, convection attribution is nearly exact in variance and mean.
Cloud attribution remains good but less precise, because the cloud contribution is smaller and its error--tendency response is less monotonic.
Attribution for the non-linear case is slightly stronger than the linear case---possibly because the richer mapping gives branch outputs more sensitivity to process-dependent differences in~$dQ$.

\paragraph{Regime-dependent errors.}
The model may fail when the coefficients themselves depend on the
state or time---for example, errors with a diurnal cycle, or errors that switch
between convective regimes or across the freezing level.  In such cases the error is no longer a function of $dQ_p$ alone.
Learning such dependence by brute force from $\sim$10{,}000 samples is
unlikely to generalise well.  Enlarging the dataset is not a satisfactory remedy,
because it would rule out the field-campaign applications that motivate
training on single steps: one month of single-site observations at 5-minute
resolution yields only $30\times24\times60/5\approx 8{,}600$ samples.

A promising route is to encode the expected structure in the function
space searched by each branch, for instance by conditioning branch outputs on
local solar time or on a regime indicator, analogous to the time-dependent
additive components of Neural Additive Time-series Models
\cite{Jo.Kim2023}.  This reinforces a broader point: PANN generalises from
small datasets only when its structure reflects expert's physical understanding of the form and complexity of the errors being sought.

\clearpage
\printbibliography

\end{document}